\documentclass[letterpaper, 10 pt, conference]{ieeeconf}  

\IEEEoverridecommandlockouts                              
\usepackage{graphics} 
\usepackage{epsfig} 
\usepackage{mathptmx} 
\usepackage{times} 
\usepackage{amsmath} 
\usepackage{amssymb}  
\usepackage{amsthm}

\title{\LARGE \bf
Inefficient Alliance Formation in Coalitional Blotto Games
}

\author{Vade Shah, Keith Paarporn, and Jason R. Marden 
\thanks{This work is supported by ONR grant \#N00014-20-1-2359 and AFOSR grants \#FA9550-20-1-0054 and \#FA9550-21-1-0203 and NSF grant  \#ECCS-2346791.}%
\thanks{V. Shah ({\tt\small vade@ucsb.edu}) and J. R. Marden are with the Department of Electrical and Computer Engineering at the University of California, Santa Barbara, CA.}%
\thanks{K. Paarporn ({\tt\small kpaarpor@uccs.edu}) is with the Department of Computer Science at the University of Colorado, Colorado Springs, CO.}
}

\theoremstyle{plain}
\newtheorem{theorem}{Theorem}

\DeclareMathOperator*{\argmax}{arg\,max}
\DeclareMathOperator*{\argmin}{arg\,min}

\begin{document}

\maketitle
\thispagestyle{empty}
\pagestyle{empty}

\begin{abstract}

When multiple agents are engaged in a network of conflict, some can advance their competitive positions by forming alliances with each other. However, the costs associated with establishing an alliance may outweigh the potential benefits. This study investigates costly alliance formation in the framework of coalitional Blotto games, in which two players compete separately against a common adversary, and are able to collude by exchanging resources with one another. Previous work has shown that both players in the alliance can mutually benefit if one player unilaterally donates, or transfers, a portion of their budget to the other. In this letter, we consider a variation where the transfer of resources is inherently inefficient, meaning that the recipient of the transfer only receives a fraction of the donation. Our findings reveal that even in the presence of inefficiencies, mutually beneficial transfers are still possible. More formally, our main result provides necessary and sufficient conditions for the existence of such transfers, offering insights into the robustness of alliance formation in competitive environments with resource constraints.

\end{abstract}

\section{Introduction}

In adversarial settings with multiple competitors, agents can often improve their performance by forming an alliance. Businesses ally to outperform rival products \cite{elmuti2001overview, culpan2002global, king2003complementary}, energy providers collaborate to succeed in markets \cite{van2015power, de2020cooperatives}, and nations join forces to confront mutual adversaries \cite{gibler2008international}. A fundamental underlying mechanism for each of these alliances is an exchange of resources, whether they be financial capital, electrical power, or military assets. By exchanging resources, agents can fortify weaknesses, complement strengths, and even intimidate adversaries.

However, when resources are lost in the process of an exchange, deciding whether to form an alliance becomes a more challenging problem. Moreover, inefficient exchanges are ubiquitous: international trades are limited by regulations and tariffs \cite{reid2001alliance, debaere2010tariffs}, energy transmissions suffer from lossy power lines \cite{paudel2020peer}, and sharing defense assets often incurs myriad losses \cite{smith1995alliance, alley2021budget}. In the presence of such limitations, a critical question emerges: \textbf{At what point do the costs of forming an alliance outweigh its potential benefits?}

\begin{figure}[t]
    \includegraphics[width=\linewidth]{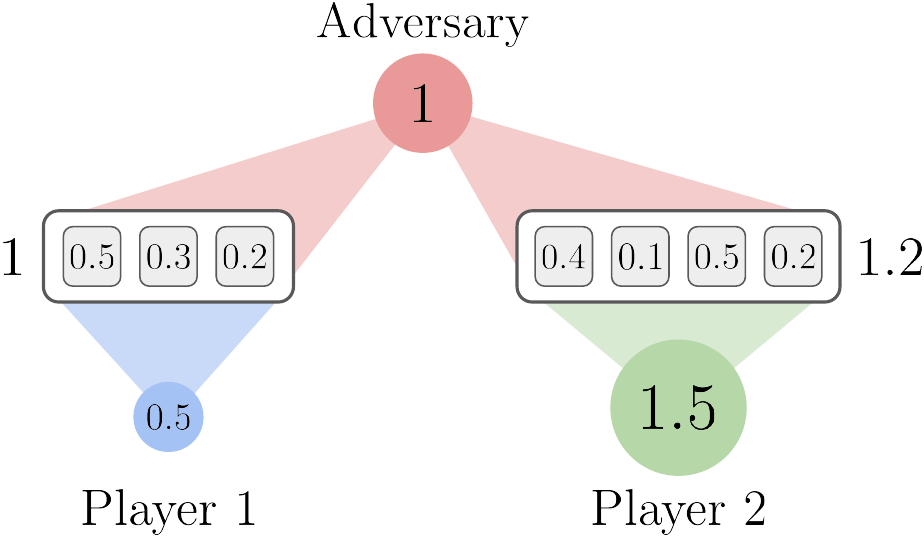}
    \caption{A cartoon depiction of a coalitional Colonel Blotto game between Players 1 and 2 and a common adversary. Player 1, Player 2, and the adversary are equipped with budgets $0.5$, $1.5$, and $1$, respectively. Player 1 and the adversary compete on the left set of contests with cumulative value $1$, and Player 2 and the adversary compete on the right set of contests with cumulative value $1.2$.}
    \centering
    \label{fig:coalitional_basic}
\end{figure}

Game theory offers a useful set of tools to study the value of alliances in a variety of adversarial contexts \cite{shenoy1979coalition, smith1995alliance, sandler1999alliance, ray2015coalition}. This work focuses on \emph{competitive resource allocation} settings like the ones mentioned above, where agents vie for prizes by strategically distributing their resources. The \emph{Colonel Blotto} game \cite{borel1921theorie, roberson2006colonel, kovenock2021generalizations} is a popular model of competitive resource allocation in which two agents compete by allocating their limited resources towards valued contests. An agent wins a contest's valuation if they allocate a greater level of resources to it than their opponent, and each agent plays with the goal of maximizing their accrued valuations.

To analyze opportunities for collaboration, we utilize the model of the \emph{coalitional} Colonel Blotto game, which has been studied extensively in the context of alliance formation \cite{kovenock2012coalitional, gupta2014three, chandan2022art, shah2024battlefield}. In the coalitional Blotto game (Figure \ref{fig:coalitional_basic}), two players compete in separate standard Blotto games against a common adversary. Previous work on these games reveals a surprising opportunity for collaboration between the two players: In certain games, if one player donates, or \emph{transfers}, a portion of their budget to the other, then both players win more contest valuation in their separate competitions than they would have had no transfer occurred \cite{kovenock2012coalitional}. That is, one player becomes \lq weaker\rq, the other becomes \lq stronger\rq, but they \emph{both} do better because of the transfer. Transferring resources causes the adversary to alter their allocation strategy, so the players can perform a transfer that manipulates the adversary to their advantage.

However, if resources are lost in the process of an exchange, it is unclear whether this strategy remains viable. In fact, it is known that in the extreme case where transfers are lost entirely (i.e., one player simply disposes of some portion of their budget without the other player receiving anything), it is impossible for both players to improve simultaneously \cite{chandan2022art}. In this work, we probe the granularity of this result to better understand the limiting effects of costly exchange. In particular, we study the case in which one player donates a portion of their budget, but the other player receives only a fraction of the sent amount. We interpret this as a model of \emph{inefficient} alliance formation, where the amount that is lost in the transfer represents the inefficiencies associated with forming the alliance.  Within this context, we seek to understand how losses affect the feasibility of budget transfers as a means of \emph{mutually beneficial} alliance formation, meaning that the transfer improves the equilibrium payoff associated with both players.


\begin{figure*}
    \includegraphics[width=\textwidth]{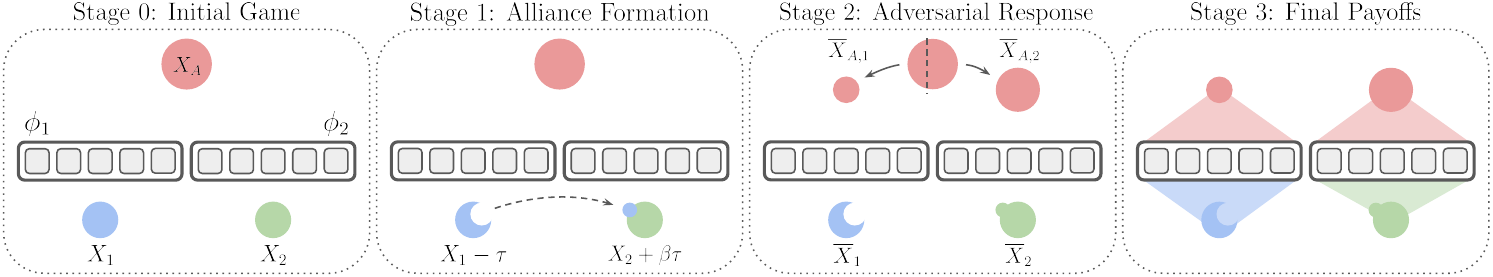}
    \caption{The stages of the coalitional Blotto game. In Stage 0, the game is initialized. In Stage 1, the two players perform mutually beneficial transfers. In Stage 2, the adversary decides how to divide their budget. In Stage 3, the two disjoint Blotto games are played.}
    \label{fig:stages}
\end{figure*}


In our first contribution, summarized in Theorem \ref{thm:mutually_beneficial}, we demonstrate that so long as the recipient receives a positive fraction of the transfer, there exists a mutually beneficial budget transfer in a nontrivial subset of games. In particular, \textbf{we provide necessary and sufficient conditions for when forming an alliance is mutually beneficial.}

Then, we abandon the requirement that transfers must benefit each player individually, and instead study the more general setting where the players seek to maximize their combined performance. In the absence of inefficiencies, it is known that the alliance can almost always improve its joint payoff by performing a transfer \cite{kovenock2012coalitional}. Our second result asserts that this conclusion no longer holds in the presence of inefficiencies. Specifically, Theorem \ref{thm:coalition_optimal} asserts that \textbf{inefficiencies eliminate opportunities for the alliance to improve its outcome in a nontrivial subset of games}. Finally, we summarize these results in elucidating simulations.

\section{Model}

\subsection{Colonel Blotto Game}

We begin our technical discussion with the classical Colonel Blotto game, where two agents (say, Player 1 and the adversary) simultaneously compete across a set of $n$ contests with valuations $v^1, \dots, v^n \geq 0$. Player 1 and the adversary are endowed with budgets $X_1 \in \mathbb{R}$ and $X_A \in \mathbb{R}$, respectively, and must decide how to allocate their resources to contests. We denote a valid allocation decision for Player 1 (and similarly for the adversary) by the tuple $d_1 = (d_1^1, \dots, d_1^n)$ where $d_1^k \geq 0$ and $\sum_{k = 1}^n d_1^k \leq X_1$. Player 1's payoff for a given allocation decision $(d_1, d_A)$ is of the form
\begin{equation}
    U_1(d_1, d_A) = \sum_{k=1}^n v^k \cdot I \{ d_1^k \geq d_A^k\},
\end{equation}
where $I\{\cdot\}$ is the usual indicator function, and the adversary's payoff is $U_A(d_1, d_A) = \sum_{k=1}^n v^k - U_1(d_1, d_A)$.

Despite the apparent simplicity of the model, characterizing Nash equilibrium allocation decisions in the classical Colonel Blotto game is a historically challenging endeavor that remains an open problem. Thus, in this work, we focus instead on the General Lotto formulation, a popular variant of the Colonel Blotto game that admits more tractable solutions by requiring agents' allocation decisions to be less than their actual budgets only \emph{in expectation}. The equilibrium payoffs\footnote{Here, we present results purely regarding the equilibrium payoffs and direct the reader to \cite{kovenock2021generalizations} for more details regarding the equilibrium strategies themselves.} for General Lotto games have been characterized in \cite{kovenock2021generalizations} and are given by
\begin{equation*}
    U_1^{\rm NE} = \begin{cases}
        \phi \left( \frac{X_1}{2 X_A} \right) & X_1 \leq X_A \\
        \phi \left( 1 - \frac{X_A}{2 X_1} \right) & X_1 > X_A
    \end{cases}
    \quad \text{ and } \quad U_A^{\rm NE} = \phi - U_1^{\rm NE},
\end{equation*}
where $\phi = \sum_{i = 1}^n v_i$ is the cumulative valuation of all of the contests in the game. We occasionally omit the dependence on the parameters $(X_1, X_A, \phi)$, i.e., we express the equilibrium payoffs as $U_1^{\rm NE}$ instead of $U_1^{\rm NE}(X_1, X_A, \phi)$, for brevity when the dependence is clear.

\subsection{Inefficient Coalitional Colonel Blotto Game}

We now turn our attention to the coalitional Colonel Blotto game, depicted in Figure \ref{fig:stages}, where two independent and self-interested players compete against across disjoint sets of contests. Here, we seek to identify whether mutually beneficial alliances exist between these two players, i.e., alliances that improve the performance of both players. Informally, an alliance involves the transfer of budgetary resources from one player to another. Our analysis will specifically focus on the \emph{inefficiencies} associated with alliance formation, meaning that these transfers are accompanied with particular losses. The coalitional Colonel Blotto game proceeds as follows.

\vspace{.1cm}

\subsubsection*{Stage 0: Initial Game} The game is initialized (Figure \ref{fig:stages}, left). Two players, referred to as Players 1 and 2, participate in standard Colonel Blotto games 1 and 2, respectively, against a common adversary who competes in both games. Player $i \in \{1, 2\}$ uses their budget $X_i$ to compete in Colonel Blotto game $i$ for valued contests with total valuation $\phi_i$. The adversary is equipped with a normalized budget $X_A = 1$, so that a coalitional Colonel Blotto game $G$ is fully parameterized by $G = (\phi_1, \phi_2, X_1, X_2) \in \mathbf{G} = \mathbb{R}_{>0}^4$.

\vspace{.1cm}

\subsubsection*{Stage 1: Alliance Formation} The players consider the formation of an alliance which allows for the transfer of budget from one player to the other.

\vspace{.1cm}

\noindent \emph{-- Transfer}: 
We denote a transfer by $\tau \in (-X_2, X_1)$, which represents the net amount of budget sent from Player 1 to Player 2. Here, a negative value indicates that the transfer goes in the opposite direction.

\vspace{.1cm}

A transfer effectively impacts the state of the game. More specifically, for a transfer $\tau$, the post-transfer budgets associated with each player are given by
\begin{gather*}
    \overline{X}_1 \triangleq \begin{cases}
        X_1 - |\tau| & \tau > 0 \\
        X_1 + \beta |\tau| & \tau \leq 0,
    \end{cases}
    \qquad
    \overline{X}_2 \triangleq \begin{cases}
        X_2 + \beta |\tau| & \tau > 0 \\
        X_2 - |\tau| & \tau \leq 0,
    \end{cases}
\end{gather*}
where $\beta \in (0, 1]$ captures the inefficiencies associated with transferring resources; the case $\beta = 1$ is the fully efficient setting considered in \cite{kovenock2012coalitional}. We say that a transfer $\tau$ \emph{induces} a new game $\overline{G} = (\phi_1, \phi_2, \overline{X}_1, \overline{X}_2)$.

\vspace{.1cm}

\subsubsection*{Stage 2: Adversarial Response} After having observed any transfer, the adversary decides how to divide their budget between the two standard Blotto games (Figure \ref{fig:stages}, center right). Observe that, depending the parameters of the game, the adversary can optimize their performance by strategically diverting more resources against one player. In particular, given the constant sum nature of the game, they can maximize their payoff by solving
\begin{equation}
    \argmin_{\underset{X_{A,1}+ X_{A,2} \leq X_A}{X_{A,1}, X_{A,2} \geq 0}} U_1^{\rm NE}(\overline{X}_1, X_{A,1}, \phi_1) - U_2^{\rm NE}(\overline{X}_2, X_{A,2}, \phi_2).
\end{equation}
The solution to this optimization problem is derived in \cite{kovenock2012coalitional} and summarized in the forthcoming Table \ref{table1}. We denote the adversary's optimal division by $\overline{X}_{A,1}$ and $\overline{X}_{A,2}$.

\vspace{.1cm}

\subsubsection*{Stage 3: Final Payoffs}
In the third and final stage, the two separate Colonel Blotto games are played, and the agents' equilibrium payoffs are realized (Figure \ref{fig:stages}, right). The payoffs to Players 1 and 2 are given by $U_1^{\rm NE}(\overline{X}_1, \overline{X}_{A,1}, \phi_1)$ and $U_2^{\rm NE}(\overline{X}_2, \overline{X}_{A,2}, \phi_2)$, respectively, while the adversary's payoff is given by $U_A^{\rm NE} = \phi_1 + \phi_2 - U_1^{\rm NE}(\overline{X}_1, \overline{X}_{A,1}, \phi_1) - U_2^{\rm NE}(\overline{X}_2, \overline{X}_{A,2}, \phi_2)$.

\subsection{Mutually Beneficial Alliances}

\begin{figure}
    \includegraphics[width=\linewidth]{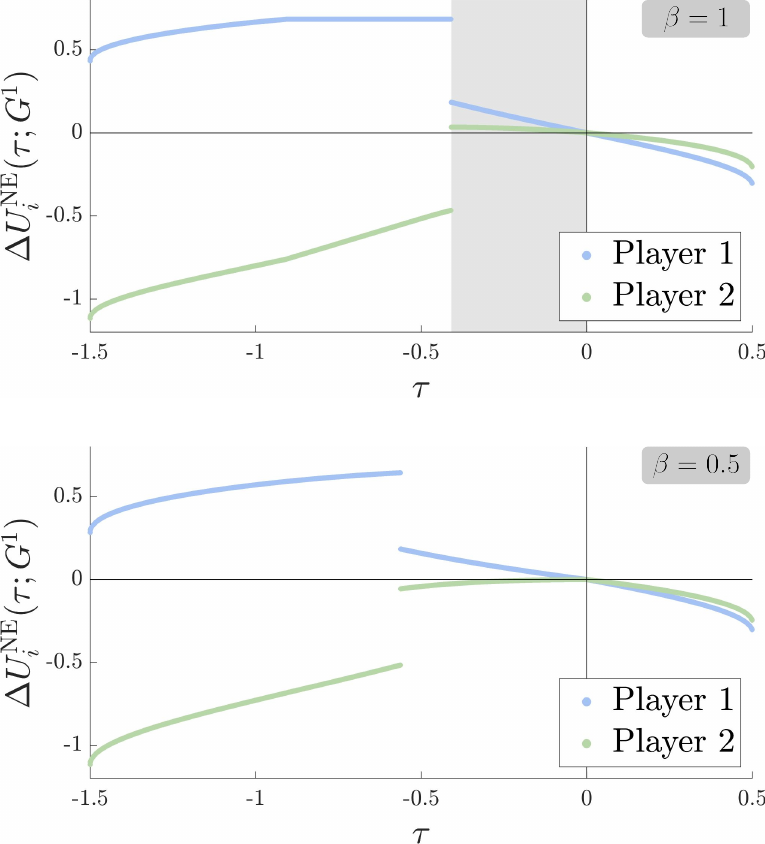}
    \caption{The change in payoff $\Delta U_i^{\rm NE}(\tau; G) \triangleq U_i^{\rm NE}(\tau; G) - U_i^{\rm NE}(\tau; 0)$ of Player $i \in \{1, 2 \}$ as a function of the budget transfer amount $\tau$ for the game $G^1 \triangleq (1, 1.2, 0.5, 1.5)$ shown in Figure \ref{fig:coalitional_basic}. When $\beta = 1$ (top), both players' payoffs increase for a range of budget transfers (grey shaded region). When $\beta = 0.5$ (bottom), both players' payoffs never increase simultaneously.}
    \label{fig:examples}
\end{figure}

The focus of this work is on identifying whether there are mutually beneficial alliances in inefficient coalitional Colonel Blotto games. Observe that the only decision for the players to make in the above game is that of the transfer $\tau$. Hence, we express the equilibrium payoffs to Players $1$ and $2$ as $U_1^{\rm NE}(\tau;G)$ and 
$U_2^{\rm NE}(\tau;G)$ to specifically highlight this dependence. We say that the players have a \emph{mutually beneficial transfer} in a game $G$ if there exists a transfer $\tau^*$ such that
$$U_1^{\rm NE}(\tau^*; G) > U_1^{\rm NE}(0; G) \ \text{and} \ U_2^{\rm NE}(\tau^*; G) > U_2^{\rm NE}(0; G),$$
meaning that both players are better off after the transfer. Figure \ref{fig:examples} provides an illustrative example demonstrating that both the donor and the recipient of an inefficient transfer can benefit in certain games. Our forthcoming results generalize this example and identify all games where such mutually beneficial alliances exist in the presence of inefficiencies. 

\section{Mutually Beneficial Inefficient Transfers}

In this section, we begin by identifying all games for which mutually beneficial budgetary transfers exist for all values of $\beta \in (0, 1]$. The forthcoming analysis will only consider games where $\frac{\phi_2}{\phi_1} \leq \frac{X_2}{X_1}$ to simplify the presentation of the results. The alternate case follows trivially by swapping the indices.

\begin{theorem}\label{thm:mutually_beneficial}
    Let $G = (\phi_1, \phi_2, X_1, X_2)$ be a coalitional Blotto game with inefficiency parameter $\beta \in (0, 1]$ where $\frac{\phi_2}{\phi_1} \leq \frac{X_2}{X_1}$. Then, $G$ has a mutually beneficial budget transfer if and only if
    \begin{gather}
        \phi_1 X_1 X_2 > \phi_2 \quad \text{ and  } \label{eq:budget_condition_1} \\
        \beta > \min \left( \left( \frac{4 \phi_2 X_1}{\phi_1 X_2^3} \right)^\frac{1}{2} - \frac{X_1}{X_2}, \left( \frac{4 \phi_2 X_1}{\phi_1 X_2} \right)^{\frac{1}{2}} + \frac{X_1}{X_2} \right). \label{eq:budget_condition_2}
    \end{gather}
    Furthermore, any mutual beneficial transfer satisfies $\tau < 0$.
\end{theorem}

\begin{figure}
    \includegraphics[width=\linewidth]{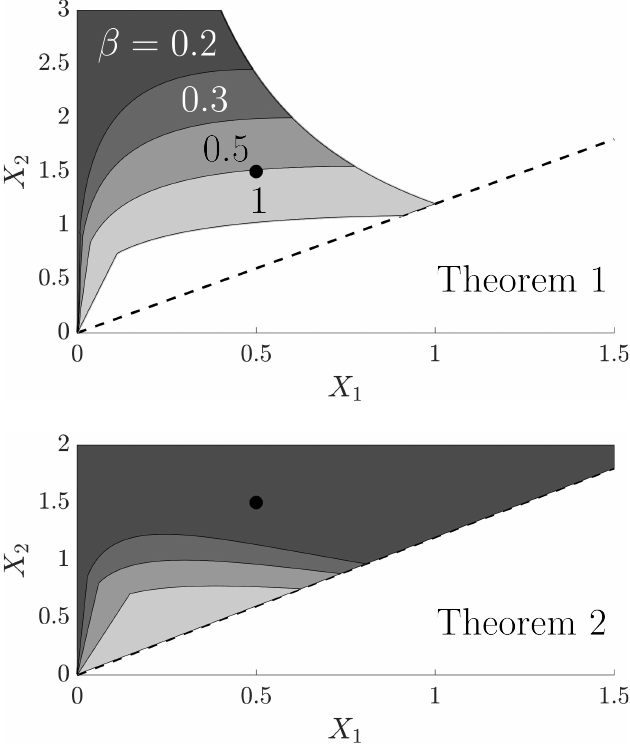}
    \caption{When $\phi_1 = 1.2$ and $\phi_2 = 1$, plots of the subsets of the parameter space in which mutually beneficial budget transfers exist (top) and alliance optimal transfers are nonzero (bottom) for various values of $\beta$ (the colors and their corresponding values of $\beta$ are the same in each plot). The horizontal and vertical axes represent $X_1$ and $X_2$, respectively, and the black dot represents $G^1$. The shading indicates layers of regions of existence (e.g., when $\beta = 1$, transfers exist across all of the shaded regions). Only games where $\frac{\phi_2}{\phi_1} \leq \frac{X_2}{X_1}$ are depicted to avoid redundancy.}
    \centering
    \label{fig:regions}
\end{figure}

Before proceeding to the proof of these statements, we make some comments. First, observe that Theorem \ref{thm:mutually_beneficial} establishes that budget transfers are viable for a nontrivial subset of games so long as the inefficiency parameter $\beta$ is positive. This is perhaps surprising, as one might expect that beyond a certain point, inefficiencies would unequivocally prohibit alliance formation, but this is not the case as shown in Figure \ref{fig:regions}. From a implementation perspective, Theorem \ref{thm:mutually_beneficial} offers a straightforward tool for players to determine whether forming an alliance is worthwhile; they can simply verify Conditions \eqref{eq:budget_condition_1} and \eqref{eq:budget_condition_2}. If they choose to investigate further, they need only check whether $\frac{\phi_2}{\phi_1} \leq \frac{X_2}{X_1}$ to decide which player should donate their budget.

\begin{proof}

The proof proceeds by analyzing each player's change in payoff as a function of the transfer $\tau$. Recall that Player $i$'s payoff is a function of their own budget as well as the adversary's optimal allocation $\overline{X}_{A,i}$ toward their standard Blotto game. The adversary's optimal allocation strategy is derived in \cite{kovenock2012coalitional} and summarized below in Table \ref{table1}. Note that when they play optimally, the adversary always uses the entirety of their budget (i.e., $\overline{X}_{A,1} + \overline{X}_{A,2} = 1$).

\begin{table}
    \centering
    \caption{The adversary's optimal budget allocation when $\frac{\phi_2}{\phi_1} \leq \frac{X_2}{X_1}$}
    \label{table1}
    \begin{center}
    \begin{tabular}{c|c|c}
        \multicolumn{1}{c|}{Case} & \multicolumn{1}{c|}{Condition} & \multicolumn{1}{c}{$\overline{X}_{A,1}$} \\
        \hline
        \vspace{0.1pt} & \vspace{0.1pt} & \vspace{0.1pt} \\
        1 & $\frac{\phi_2}{\phi_1} \neq \frac{\overline{X}_2}{\overline{X}_1} \text{ and } \frac{\phi_2}{\phi_1} \leq \overline{X}_1 \overline{X}_2$ & $1$ \\
        \vspace{0.1pt} & \vspace{0.1pt} & \vspace{0.1pt} \\
        2 & $0 < 1 - \left( \frac{\phi_1 \overline{X}_1 \overline{X}_2}{\phi_2} \right) ^{\frac{1}{2}} \leq \overline{X}_2$ & $\left( \frac{\phi_1 \overline{X}_1 \overline{X}_2}{\phi_2} \right) ^{\frac{1}{2}}$ \\
        \vspace{0.1pt} & \vspace{0.1pt} & \vspace{0.1pt} \\
        3 & $1 - \left( \frac{\phi_1 \overline{X}_1 \overline{X}_2}{\phi_2} \right) ^{\frac{1}{2}} > \overline{X}_2$ & $\frac{\left( \phi_1 \overline{X}_1 \right)^{\frac{1}{2}}}{\left( \phi_1 \overline{X}_1 \right)^{\frac{1}{2}} + \left( \phi_2 \overline{X}_2 \right)^{\frac{1}{2}}}$ \\
        \vspace{0.1pt} & \vspace{0.1pt} & \vspace{0.1pt} \\
        4 & $\frac{\phi_2}{\phi_1} = \frac{\overline{X}_2}{\overline{X}_1} \text{ and } 1 \leq \overline{X}_1 + \overline{X}_2$ & $X_{A, i} \leq X_i$
    \end{tabular}
    \end{center}
\end{table}

For ease of notation, we define the subsets
\begin{equation*}
    \mathbf{C}_i \triangleq \left\{ G \in \mathbf{G} \ \Bigg \vert \ \frac{\phi_2}{\phi_1} \leq \frac{X_2}{X_1} \text{ and } G \text{ belongs to Case $i$} \right\}
\end{equation*}
for $i \in \{ 1, 2, 3, 4 \}$. When we write that $G = (\phi_1, \phi_2, X_1, X_2) \in \mathbf{C}_i$, the reader should read this as \lq the parameters $(\phi_1, \phi_2, X_1, X_2)$ satisfy the conditions for Case $i$ in Table \ref{table1}\rq, with $X_1$ and $X_2$ in place of $\overline{X}_1$ and $\overline{X}_2$, respectively.

\subsubsection*{Case 1} Consider any game $G = (\phi_1, \phi_2, X_1, X_2) \in \mathbf{C}_1$. In this case, $\overline{X}_{A,1} = 1$, meaning that the adversary does not allocate any of their budget against Player 2. Thus, Player 2 wins all of their contests to begin with, so there is no budget transfer that can improve their payoff.

\subsubsection*{Case 2} First, consider transfers going from Player 1 to Player 2, i.e., positive transfers. If $\tau > 0$, then the induced $\overline{G} = (\phi_1, \phi_2, \overline{X}_1, \overline{X}_2)$ belongs to $\mathbf{C}_2$, so Player 1's payoff is given by $U_1^{\rm NE} (\tau; G) = \frac{1}{2} \left(\frac{\phi_1 \phi_2 (X_1 - \tau)}{X_2 + \beta \tau} \right)^{\frac{1}{2}}$. However, since $\frac{d}{d \tau} U_1^{\rm NE} < 0$ (i.e., Player 1's payoff decreases), Player 1 can never benefit by transferring their budget to Player 2.

Now, consider transfers where $\tau < 0$. A sufficiently large transfer may induce $\overline{G} \in \mathbf{C}_3$, or may result in $\frac{\phi_2}{\phi_1} \geq \frac{\overline{X}_2}{\overline{X}_1}$, but in either scenario, Player 2 would be worse off since they would be drawing the adversary's attention while losing budget. Thus, we can limit our attention to transfers that induce $\overline{G} \in \mathbf{C}_2$, in which case the players' payoffs are given by
\begin{align*}
    U_1^{\rm NE}(\tau; G) &= \frac{1}{2} \left( \frac{\phi_1 \phi_2 (X_1 - \beta \tau)}{X_2 + \tau} \right)^{\frac{1}{2}}, \\
    U_2^{\rm NE}(\tau; G) &= \phi_2 \left( 1 - \frac{1}{2 (X_2 + \tau)} \right) + \frac{1}{2} \left( \frac{\phi_1 \phi_2 (X_1 - \beta \tau)}{X_2 + \tau} \right)^{\frac{1}{2}}.
\end{align*}
Since $\frac{d}{d \tau} U_1^{\rm NE} < 0$ (i.e., Player 1's payoff increases), Player 1 will always accept a transfer from Player 2. Furthermore, if $\frac{d}{d \tau} U_2^{\rm NE} \big|_{\tau \to 0^-} < 0$, then Player 2 would benefit from transferring a sufficiently small amount of budget to Player 1. By simple calculus and algebraic manipulation, one can easily verify that $\frac{d}{d \tau} U_2^{\rm NE} \big|_{\tau \to 0^-} < 0 \iff \beta > \left( \frac{4 \phi_2 X_1}{\phi_1 X_2^3} \right)^\frac{1}{2} - \frac{X_1}{X_2}$.

\subsubsection*{Case 3} When $\tau > 0$, the induced $\overline{G}$ may belong to $\mathbf{C}_2$ or $\mathbf{C}_3$. If $\overline{G} \in \mathbf{C}_3$, then Player 1's payoff is given by $U_1^{\rm NE} (\tau; G) = \frac{1}{2} \phi_1 (X_1 - \tau) + \frac{1}{2} (\phi_1 \phi_2 (X_1 - \tau)(X_2 + \beta \tau))^{\frac{1}{2}}$. It is straightforward to show that $\frac{d}{d \tau} U_1^{\rm NE} \big\vert_{\tau \to 0^+} < 0 \iff 2 \left(\frac{\phi_1 X_1 X_2}{\phi_2} \right)^{\frac{1}{2}} < \beta X_1 - X_2$, but this is not true of any game where $\frac{\phi_2}{\phi_1} \leq \frac{X_2}{X_1}$. Furthermore, given that net positive transfers from Player 1 to 2 are not mutually beneficial in Case 2, it follows that there is no mutually beneficial transfer such that $\overline{G} \in \mathbf{C}_2$. Thus, any positive transfer cannot be mutually beneficial.

When $\tau < 0$, the induced $\overline{G}$ belongs to $\mathbf{C}_3$, where the players' payoffs are given by
\begin{align*}
    U_1^{\rm NE}(\tau; G) &= \frac{1}{2} \phi_1 (X_1 - \beta \tau) + \frac{1}{2} (\phi_1 \phi_2 (X_1 - \beta \tau)(X_2 + \tau))^{\frac{1}{2}}, \\
    U_2^{\rm NE}(\tau; G) &= \frac{1}{2} \phi_2 (X_2 + \tau) + \frac{1}{2} (\phi_1 \phi_2 (X_1 - \beta \tau)(X_2 + \tau))^{\frac{1}{2}}.
\end{align*}
It is straightforward to show that if $\frac{d}{d \tau} U_2^{\rm NE} \big|_{\tau \to 0^-} \geq 0$, then it will remain nonnegative for all $-X_2 < \tau \leq 0$. Furthermore, since $\frac{d}{d \tau} U_2^{\rm NE} \geq \frac{d}{d \tau} U_1^{\rm NE}$, both players will strictly benefit from a net positive transfer when $\frac{d}{d \tau} U_2^{\rm NE} \big|_{\tau \to 0^-} < 0 \iff \beta > \left( \frac{4 \phi_2 X_1}{\phi_1 X_2} \right)^{\frac{1}{2}} + \frac{X_1}{X_2}$. This is also a necessary condition for a mutually beneficial transfer such that $\overline{G} \in \mathbf{C}_2$, since the subset of games that satisfy the Case 2 condition can be reached by a transfer starting from Case 3 only if this condition is satisfied.

\subsubsection*{Case 4} For any game $G \in \mathbf{C}_4$, we show in Theorem \ref{thm:coalition_optimal} that the sum of the players' payoffs is maximized, so there is no mutually beneficial transfer.

\end{proof}

\section{Alliance Optimal Inefficient Transfers}

The previous section demonstrates that in spite of inefficiencies, players can still form mutually beneficial alliances in many cases. In this section, we study how inefficiencies impact opportunities to form alliances that are not necessarily mutually beneficial, but rather jointly optimal. Mathematically speaking, we study transfers that solve
\begin{equation*}
    \argmax_{\tau} \ U_{1, 2}^{\rm NE}(\tau; G),
\end{equation*}
where $U_{1, 2}^{\rm NE}(\tau; G) \triangleq U_1^{\rm NE}(\tau; G) + U_2^{\rm NE}(\tau; G)$ is the \emph{alliance payoff}. We call a transfer $\tau^{\dagger}$ that maximizes this sum \emph{alliance optimal}. Note that if a transfer is mutually beneficial, then it must also improve the alliance payoff, but the converse need not be true. Theorem \ref{thm:coalition_optimal} effectively characterizes the gap between these two forms of strategic opportunities.

\begin{theorem}\label{thm:coalition_optimal}
    Let $G = (\phi_1, \phi_2, X_1, X_2)$ be a coalitional Blotto game with inefficiency parameter $\beta \in (0, 1]$ where $\frac{\phi_2}{\phi_1} \leq \frac{X_2}{X_1}$. Then, the alliance optimal transfer $\tau^{\dagger}$ is nonzero if and only if $G \in \mathbf{G} \setminus \mathbf{G}^{\dagger}(\beta)$, where
    \begin{equation*}
        \begin{split}
            \mathbf{G}^{\dagger}(\beta) \triangleq \ &\left\{ G \in \mathbf{C}_2 \ \Bigg\vert \ X_1 + \beta X_2 \leq \left( \frac{\phi_2 X_1}{\phi_1 X_2} \right)^{\frac{1}{2}} \right\} \ \cup \ \mathbf{C}_4 \ \cup \\
            &\left\{ G \in \mathbf{C}_3 \ \Bigg\vert \ \beta \phi_1 - \phi_2 \leq \left( \frac{\phi_1 \phi_2}{X_1 X_2} \right)^{\frac{1}{2}} (X_1 - \beta X_2) \right\}.
        \end{split}
    \end{equation*}
\end{theorem}

It is straightforward to verify that, as one might expect, the set of games in which there exist mutually beneficial transfers is a subset of the set of games in which there exist nonzero alliance optimal transfers. This result is depicted graphically in Figure \ref{fig:regions}. However, there is an interesting technical distinction between these two regions regarding their dependence on the inefficiency parameter $\beta$: Although the players cannot always mutually improve even in efficient environments, the alliance can almost always improve only in efficient environments. That is, the set of games in which mutually beneficial transfers do not exist has positive measure for all values of $\beta$, but the set of games in which alliance optimal transfers are zero has positive measure only when $\beta < 1$; when $\beta = 1$, one can easily verify that $\mathbf{G}^{\dagger}(1)$ is the measure-zero subset of $G$ where $\frac{\phi_2}{\phi_1} = \frac{X_2}{X_1}$. In this measure-theoretic sense, inefficiencies have a more pronounced impact on the outcome of the alliance than they do on the individual.

\begin{proof}

The proof proceeds by analyzing the alliance payoff in each of the four cases and computing its derivative.

\subsubsection*{Case 1} In this case, since the adversary allocates all of their budget towards Player 1, the players can improve the sum of their payoffs by transferring $\tau < 0$ from Player 2 to Player 1 until the induced $\overline{G}$ satisfies either Case 2 (in which case they proceed according to Case 2 below) or Case 4.

\subsubsection*{Case 2} When $\tau > 0$, the alliance payoff is given by
\begin{equation*}
    U_{1, 2}^{\rm NE}(\tau; G) = \phi_2 \left( 1 - \frac{1}{2 (X_2 + \beta \tau)} \right) + \frac{1}{2} \left( \frac{\phi_1 \phi_2 (X_1 - \tau)}{X_2 + \beta \tau} \right)^{\frac{1}{2}},
\end{equation*}
and its derivative $\frac{d}{d \tau} U_{1, 2}^{\rm NE}$ is positive when
\begin{equation*}
    X_1 + \frac{1}{\beta} X_2 < \left( \frac{\phi_2 (X_1 - \tau)}{\phi_1 (X_2 + \beta \tau)} \right)^{\frac{1}{2}}.
\end{equation*}
However, it is relatively straightforward to show that this condition is not satisfied for any game $\mathbf{C}_2$, so it follows that any positive transfer cannot cause an increase in $U_{1, 2}^{\rm NE}(\tau; G)$. When $\tau < 0$, the alliance payoff is given by
\begin{equation*}
    U_{1, 2}^{\rm NE}(\tau; G) = \phi_2 \left( 1 - \frac{1}{2 (X_2 + \tau)} \right) + \frac{1}{2} \left( \frac{\phi_1 \phi_2 (X_1 - \beta \tau)}{X_2 + \tau} \right)^{\frac{1}{2}}.
\end{equation*}
The derivative $\frac{d}{d \tau} U_{1, 2}^{\rm NE}$ remains negative so long as $$\overline{X}_1 + \beta \overline{X}_2 > \left( \frac{\phi_2 \overline{X}_1}{\phi_1 \overline{X}_2} \right)^{\frac{1}{2}}.$$ Thus, in any game where $\frac{d}{d \tau} U_{1, 2}^{\rm NE} \big\vert_{\tau \to 0^-} < 0$, Player 2 should transfer budget to Player 1 until either $\overline{G} \in \mathbf{C}_1$ (in which case they proceed according to Case 1 above) or $\overline{G} \in \mathbf{C}_4$, or until $\overline{X}_1 + \beta \overline{X}_2 = \left( \frac{\phi_2 \overline{X}_1}{\phi_1 \overline{X}_2} \right)^{\frac{1}{2}}$. In the latter case, the alliance payoff is maximized, and $\tau^{\dagger}$ is the unique transfer that satisfies this condition. Otherwise, if $\frac{d}{d \tau} U_{1, 2}^{\rm NE} \big\vert_{\tau \to 0^-} > 0$, then the payoff to the alliance is already maximized and $\tau^{\dagger} = 0$.

\subsubsection*{Case 3} Following a similar procedure as above, it is straightforward to show that when $\tau > 0$, $\frac{d}{d \tau} U_{1, 2}^{\rm NE} < 0$ for all games in $\mathbf{C}_3$; thus, any positive transfer cannot improve the alliance payoff. When $\tau < 0$, $U_{1, 2}^{\rm NE}(\tau; G)$ is given by
\begin{align*}
    U_{1, 2}^{\rm NE}(\tau; G) = &\frac{1}{2} \phi_1 (X_1 - \beta \tau) + \frac{1}{2} \phi_2 (X_1 + \tau) \\
    + &(\phi_1 \phi_2 (X_1 - \beta \tau) (X_2 + \tau))^{\frac{1}{2}}.
\end{align*}
The derivative $\frac{d}{d \tau} U_{1, 2}^{\rm NE}$ remains negative so long as $$\beta \phi_1 - \phi_2 > \left( \frac{\phi_1 \phi_2}{\overline{X}_1 \overline{X}_2} \right)^{\frac{1}{2}} (\overline{X}_1 - \beta \overline{X}_2).$$ Thus, in any game where $\frac{d}{d \tau} U_{1, 2}^{\rm NE} \big\vert_{\tau \to 0^-} < 0$, Player 2 should transfer budget to Player 1 until $\overline{G} \in \mathbf{C}_2$ (in which case they proceed according to Case 2 above), or until $\beta \phi_1 - \phi_2 = \left( \frac{\phi_1 \phi_2}{\overline{X}_1 \overline{X}_2} \right)^{\frac{1}{2}} (\overline{X}_1 - \beta \overline{X}_2)$, at which point the alliance payoff is maximized. Otherwise, if $\frac{d}{d \tau} U_{1, 2}^{\rm NE} \big\vert_{\tau \to 0^-} > 0$, then the payoff to the alliance is already maximized and $\tau^{\dagger} = 0$.

\subsubsection*{Case 4} In Cases 1 and 2, the alliance payoff increases when Player 2 transfers budget to Player 1 until reaching a maximum, or until reaching Case 4. The symmetrical statement is true for all games where $\frac{\phi_2}{\phi_1} \geq \frac{X_2}{X_1}$, meaning that every coalition-improving transfer moves towards Case 4. Since $U_{1,2}^{\rm NE}$ is continuous, we can conclude that if $G$ belongs to Case 4, then $U_{1, 2}^{\rm NE}$ is already at a maximum, and $\tau^{\dagger} = 0$; similarly, if $\overline{G}$ belongs to Case 4, then $U_{1, 2}^{\rm NE}$ is also at a maximum, and $\tau^{\dagger}$ is the transfer that induces $\overline{G}$.

\end{proof}

\begin{figure}
    \includegraphics[width=\linewidth]{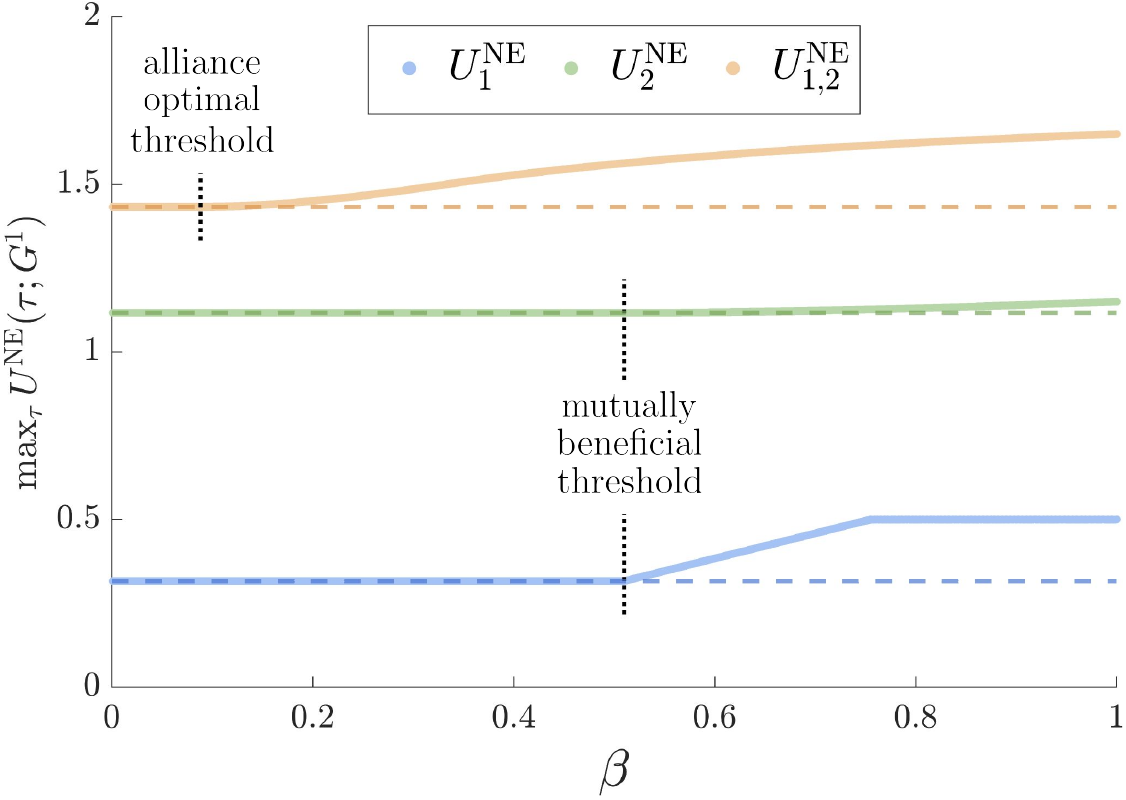}
    \caption{The maximum values of $U_1^{\rm NE}$ (blue), $U_2^{\rm NE}$ (green), and $U_{1, 2}^{\rm NE}$ (orange) for the game $G^1$ (Figure \ref{fig:coalitional_basic}) for various values of $\beta$. The colored horizontal dashed lines depict the corresponding nominal value when no transfer occurs. The black vertical dashed line on the top left indicates the threshold for the existence of nonzero coalition optimal transfers; the one in the center indicates the threshold for the existence of mutually benefical transfers. Note that the maximizing transfer is chosen separately for each curve, subject to its respective constraints.}
    \label{fig:sim}
\end{figure}

We conclude our discussion by comparing the two forms of strategic alliance in simulation. Figure \ref{fig:sim} shows the maximum possible values of $U_1^{\rm NE}$, $U_2^{\rm NE}$, and $U_{1, 2}^{\rm NE}$ for the game $G^1$ with various values of $\beta$. Players can only perform mutually beneficial transfers when $\beta \gtrapprox 0.5099$, meaning that about half of the transfer must be received in order for players to benefit. In contrast, Theorem \ref{thm:coalition_optimal} identifies the threshold $\beta \approx 0.0883$ for alliance optimal transfers, meaning that even if only 10\% of the exchange is completed, performing a transfer can still improve their joint payoff.

\section{Conclusion}

In this work, we examine a three-stage coalitional Colonel Blotto game in which two players compete against a common adversary by allocating their limited budgets towards valued contests. We first show that under certain conditions, players can form mutually beneficial alliances by transferring their budgets even in the presence of inefficiencies. Then, we study the alliance's optimal performance, and we demonstrate that inefficiencies limit opportunities for mutual improvement in a nontrivial subset of games. These results lend novel insight into the effects of inefficiencies on alliance formation and prompt further investigation using practical examples.

\bibliographystyle{ieeetr}
\bibliography{references}

\end{document}